\documentclass[12pt,oneside,letterpaper]{article}
\usepackage{titlesec}
\titleformat*{\section}{\large\bfseries}
\titleformat*{\subsection}{\normalsize\bfseries}
\usepackage{amssymb}
\usepackage{amsmath}
\usepackage[dvips]{graphicx}
\usepackage{setspace}
\usepackage{amsfonts}
\usepackage{fancyhdr}
\usepackage{xcolor}
\usepackage{caption}
\usepackage{graphicx}
\usepackage{rotating}
\usepackage{comment}
\usepackage{color}
\usepackage{cite}
\usepackage{braket}
\definecolor{darkgreen}{rgb}{0,0.5,0}
\definecolor{darkblue}{rgb}{0,0,0.6}
\definecolor{purple}{rgb}{0.4,.2,0.7}
\newcommand{\p}{\partial}

\newcommand{\f}{\frac}

\newcommand{\be}{\begin{equation}}
\newcommand{\ee}{\end{equation}}

\usepackage[colorlinks=true,citecolor=darkgreen,linkcolor=black,urlcolor=purple]{hyperref}

\usepackage{pdfsync}
\makeatletter
\newcommand*{\defeq}{\mathrel{\rlap{%
                     \raisebox{0.3ex}{$\m@th\cdot$}}%
                     \raisebox{-0.3ex}{$\m@th\cdot$}}%
                     =} 
\makeatother

\DeclareMathOperator{\Tr}{Tr}
\def\be{\begin{eqnarray}}
\def\ee{\end{eqnarray}}

\newcommand{\bea}{\begin{eqnarray}}
\newcommand{\eea}{\end{eqnarray}}
\newcommand{\evac}{\varepsilon_{\text{vac}}}
\def\ben{\begin{equation}}
\def\een{\end{equation}}

\let\a=\alpha \let\b=\beta \let\g=\gamma \let\d=\delta 
  \let\q=\theta 
\let\l=\lambda \let\m=\mu \let\n=\nu  \let\p=\phi \let\r=v
\let\s=\sigma \let\t=\tau

\let\w=\omega \let\G=\Gamma \let\D=\Delta

\let\f=\frac

\def\be{\begin{equation}}
\def\ee{\end{equation}}
\def\ba{\begin{array}}
\def\ea{\end{array}}

\def\vol{{\rm Vol}}

\def\ba#1\ea{\begin{align}#1\end{align}}
\def\bs#1\es{\begin{split}#1\end{split}}

\renewcommand{\p}{\partial}

\usepackage[top=1in, bottom = 1in, left = 1in, right = 1in]{geometry}

\allowdisplaybreaks  % allow page breaks in displayed eqs

\begin{document}
\onehalfspacing

\begin{center}

~
\vskip5mm

{\LARGE  {
Modular invariance and thermal\\ \vspace{5mm}  effective field theory in CFT
}}

\vskip8mm

Kuroush Allameh and Edgar Shaghoulian

\vskip8mm
{ \it UC Santa Cruz\\
Physics Department\\
1156 High Street\\
Santa Cruz, CA 95064}

%\tt{eshaghoulian@ucsc.edu}

\end{center}

\vspace{4mm}

\begin{abstract}
\noindent
 We use thermal effective field theory to derive that the coefficient of the first subleading piece of the thermal free energy, $c_1$, is equal to the coefficient of the subleading piece of the Casimir energy on $S^1 \times S^{d-2}$ for $d \geq 4$. We conjecture that this coefficient obeys a sign constraint $c_1 \geq 0$ in CFT and collect some evidence for this bound. We discuss various applications of the thermal effective field theory, including placing the CFT on different spatial backgrounds and turning on chemical potentials for $U(1)$ charge and angular momentum. Along the way, we derive the high-temperature partition function on a sphere with arbitrary angular velocities using only time dilation and length contraction. \end{abstract}

%\tt{600 Warren Road \#4-3F\\
%Ithaca, NY 14850}

%\tt{Submitted on March 31, 2020}

\thispagestyle{empty}
\pagebreak
\pagestyle{plain}

\setcounter{tocdepth}{2}
{}
\vfill
\clearpage
\setcounter{page}{1}

\tableofcontents

\section{Introduction}
Conformal field theories (CFTs) in two dimensions are famously modular invariant. This is an invariance of the thermal partition function at finite size under $SL(2,\mathbb{Z}$) transformations, which makes up the group of large diffeomorphisms of the torus. This provides many constraints on the theory, the first of which was an asymptotic formula for the high-energy density of states \cite{Cardy:1986ie}. Since then, many implications of modular invariance have been explored in two dimensions, including constraints on operator product expansion (OPE) coefficients \cite{Kraus:2016nwo, Das:2017vej, Cardy:2017qhl, Cho:2017fzo, Keller:2017iql, Das:2017cnv, Pal:2019yhz}, upper bounds on the smallest operator dimension and twist gap above the vacuum \cite{Hellerman:2009bu, Collier:2016cls, Hartman:2019pcd, Benjamin:2019stq, Pal:2022vqc}, extension of the range of validity of the Cardy formula \cite{Hartman:2014oaa, Anous:2018hjh}, and applications to holography \cite{Keller:2011xi, Benjamin:2015hsa, Belin:2016dcu, Belin:2018oza}, to name a few.

Given the plethora of applications of modular invariance, it is important to understand its role in higher dimensions. For a CFT on a torus $T^{d-1}$, the thermal partition function is similarly $SL(d,\mathbb{Z})$ invariant. Implications of this invariance (and its invariant subgroups) have been explored in \cite{Shaghoulian:2015kta, Shaghoulian:2015lcn, Belin:2016yll, Levine:2022wos, Luo:2022tqy}, in particular providing novel constraints on the Casimir energy of the CFT.\footnote{An alternative approach to providing novel constraints on the CFT from finite-temperature considerations is to consider the constraints coming from the KMS conditions \cite{El-Showk:2011yvt}, see e.g. \cite{Iliesiu:2018fao, Marchetto:2023fcw}.}

Placing the CFT on $S^{d-1}$ is a natural case to consider, since the states on $S^{d-1}$ are in one-to-one correspondence with the local operators of the theory. However, the thermal partition function of CFT on $S^{d-1}$ has no obvious modular invariance. The key thing that is needed for some sort of modular invariance is a spatial circle that can be swapped with the thermal circle. One possible approach to producing such a spatial circle is to restrict to even-dimensional CFTs on an odd-dimensional sphere written as a circle fibration. We can then take a lens space quotient of the Hopf fiber, and in the limit of large quotient we have an ``emergent circle" that we can swap with the thermal circle \cite{Shaghoulian:2016gol}:
\be\label{emergent}
p, k \gg 1: \qquad Z[S^1_{2\pi/k} \times S^3/\mathbb{Z}_p] \approx Z[S^1_{2\pi/p} \times S^3/\mathbb{Z}_k].
\ee
Another approach is to use Weyl invariance to isolate a circle factor \cite{Horowitz:2017ifu}:
\be
ds^2 = d\t_\b^2 + d\q^2 + \sin^2 \q \, d\Omega_{d-3}^2 + \cos^2 \q \, d\phi^2 \approx \f{d\t_\b^2 + d\q^2 + \sin^2 \q\, d\Omega_{d-3}^2}{\cos^2 \q} + d\phi^2 \,.
\ee
In this case we relate the thermal partition function at some inverse temperature $\b$ on $S^{d-1}$ to the thermal partition function at inverse temperature $2\pi$ on $\mathcal{H}^{d-1}/\mathbb{Z}$, where the quotient refers to the periodicity $\b$ of the $\t$ coordinate. 

In this paper we will provide an alternative approach, that of thermal effective field theory, which will let us directly address the thermal partition function on $S^{d-1}$. We will primarily be interested in the coefficient of the leading subextensive piece of the thermal effective action $c_1$. We will conjecture a sign constraint on this coefficient $c_1 \geq 0$, and provide some evidence. Modular invariance will come in through the thermal effective field theory allowing us to relate this coefficient to the subleading piece of the Casimir energy on $S^1 \times S^{d-2}$. Along the way, we will use the thermal effective field theory to derive various other properties of the CFT, for example we will provide a derivation of \eqref{emergent} and consider extensions to one-point functions and partition functions with chemical potentials turned on. The use of thermal effective field theory has seen a resurgence over the past years, beginning in this context with \cite{Bhattacharyya:2007vs, Jensen:2012jh, Banerjee:2012iz} and continuing up to the present day, with recent work using these ideas in similar contexts including \cite{DiPietro:2014bca, Horowitz:2017ifu, Kang:2022orq,  Luo:2022tqy, Benjamin:2023qsc}

In a related but somewhat tangential vein, we will provide a novel derivation of the free energy at high temperature and arbitrary angular potentials using only Lorentz invariance. This formula was first derived using the fluid approximation in \cite{Bhattacharyya:2007vs} and more recently using thermal effective field theory in \cite{Benjamin:2023qsc}. This in turn reproduces the formulas for the asymptotic density of local operators at large scaling dimension and fixed spin \cite{Shaghoulian:2015lcn}.

\section{Thermal effective field theory}
\subsection{Modular invariance at subextensive order}
In this section we combine modular invariance with some basic effective field theory to obtain a constraint on the subleading term in the free energy of a quantum field theory on $S^1_\b \times \mathcal{M}^{d-1}$. We consider a spatial manifold $\mathcal{M}^{d-1}$ with finite volume $\int d^{d-1} \Sigma\,\,$  and assume the theory is gapped at finite temperature. That means we can dimensionally reduce over the $S^1_\b$ and write an effective action\footnote{If we wanted to study perturbations around equilibrium, then we would include an effective action for the  hydrodynamic modes which are not gapped (since the stress tensor is conserved). But since we will only study equilibrium properties we are free to ignore this. See \cite{Bhattacharyya:2007vs, Jensen:2012jh, Banerjee:2012iz} for applications to hydrodynamics.}
\be
I = -\log Z(\b) = \int d^{d-1} \Sigma \left(-c_0 \f{1}{\b^{d-1}} + c_1 \f{R}{\b^{d-3}} +\cdots\right) + \text{non-pert.}
\ee
The effective action is written in terms of the background terms available to the theory, which in this case is just the Riemann curvature tensor and its contractions, along with derivatives.  We see the identity operator provides the leading, extensive piece of the free energy. The first subextensive correction is given by the Ricci scalar $R$. At quadratic and higher order there are in general distinct contractions we can write down, e.g. $R^2$ and $R_{\m\n} R^{\m\n}$. Derivatives that are not total derivatives first appear at the same order as $\b^{7-d}$, in terms like $R_{\m\n} \Box R^{\m\n}$.\footnote{For a maximally symmetric space like $S^{d-1}$ or $H^{d-1}$ all derivative terms vanish, since $R_{\m\n\rho\s} \propto g_{\m\rho}g_{\n\s} - g_{\n\rho}g_{\m\s}$ and the metric is covariantly conserved $\nabla_{\a} g_{\m\n} = 0$. Furthermore the Riemann tensor is determined in terms of the Ricci scalar, so the effective theory can be thought of as a series in powers of the Ricci scalar.} The coefficient $c_0 \geq 0$ but there are currently no constraints on $c_1$. We conjecture $c_1 \geq 0$ and collect some evidence in Section \ref{conjsec}. 

Notice that because the leading extensive piece does not care about the curvature of the manifold, we can preserve the value of the partition function at this order by replacing the spatial manifold $\mathcal{M}^{d-1}$ with any spatial manifold of the same volume. In particular we can replace it with a torus $T^{d-1}$, which has nice modular properties. We can then switch quantizations by trading a spatial cycle in $T^{d-1}$ with the thermal cycle and therefore map the high-temperature partition function into a low-temperature partition function. (This is a low-temperature partition function in the sense that at leading order in small $\b$ we pick up only the vacuum contribution in this channel, even though the inverse temperature in this channel is not the largest scale around; see Appendix A of \cite{Shaghoulian:2015kta} for a justification.) In this way we see that the leading extensive piece of the high-temperature partition function on $\mathcal{M}^{d-1}$ is related to the vacuum energy of the theory on $T^{d-1}$ with one spatial cycle taken very small. This can be modified to trade $\mathcal{M}^{d-1}$ with an $S^1 \times \widetilde{\mathcal{M}}^{d-2}$ of the same volume, since all that we needed for the preceding argument was a single spatial cycle. It is important in this argument that the leading piece of the vacuum energy on $S^1 \times \widetilde{\mathcal{M}}^{d-2}$ for small $S^1$ only depends on the volume of $\widetilde{\mathcal{M}}^{d-2}$ (and the volume of the $S^1$), not on any local characteristics of the manifold. This lets us relate the asymptotic density of states on $S^{d-1}$ (and therefore the asymptotic density of local operators) to the Casimir energy on $S^1 \times \widetilde{\mathcal{M}}^{d-2}$. 

With the effective action written above we can go further. We need to restrict to $d \geq 4$. In that case, we can replace $\mathcal{M}^{d-1}$ with a manifold $S^1 \times \widetilde{\mathcal{M}}^{d-2}$ with the same volume \emph{and} the same integrated Ricci scalar, matching the first two terms of the effective action. This is possible since we have two length scales we can vary, the size of the $S^1$ and the size of the $\widetilde{\mathcal{M}}^{d-2}$. If the integrated Ricci scalar of $\mathcal{M}^{d-1}$ is positive, then we can make the minimal choice of an $S^1 \times S^{d-2}$. If it's negative, then we can make this same choice and simply flip the sign of first subleading correction (or one can pick an $S^1 \times H^{d-2}$ for compact $H^{d-2}$). If it's zero, then the first subleading term in the effective action vanishes. 

To see how this works in an explicit example -- the one we will care about in this work -- consider $\mathcal{M}^{d-1} = S_L^{d-1}$, i.e. a sphere of radius $L$. We can replace this with $S^1_{L_1} \times S^{d-2}_{L_2}$.   We tune the two free parameters $L_1$ (the {\bf circumference} of the $S^1$) and $L_2$ (the {\bf radius} of the $S^{d-2}$),  to preserve both the overall volume of the spatial manifold \emph{and} the integrated Ricci scalar. For example, for $d=4$ we have
\begin{align}
\vol(S^3_L) = 2\pi^2 L^3\,,\qquad \text{Ricci}(S^3_L) = \f{6}{L^2}\\
\vol(S^1_{L_1} \times S^2_{L_2}) = 4\pi L_2^2 L_1 \,,\qquad \text{Ricci}(S^1_{L_1} \times S^2_{L_2}) = \f{2}{L_2^2}
\end{align}
Equating the volumes and the Ricci scalars gives  $L_1 = 3\pi L/2$ and $L_2 = L/\sqrt{3}$. (In this case since the Ricci scalars are constant, equating the integrated Ricci scalars and volumes is the same as equating the local Ricci scalars and volumes.) So we have
\be\label{firstbox}
\hspace{-12mm}\boxed{\text{to first subextensive order:\quad} Z(\b)_{S^3_L} = Z\left(\b\right)_{S^1_\f{3\pi L}{2} \times S^2_{L/\sqrt{3}}}= Z\left(\f{3\pi L}{2}\right)_{S^1_\b \times S^2_{L/\sqrt{3}}}}
\ee
At high temperature $\b \rightarrow 0$ we can write as before
\be
-\log Z(\b)_{S^3_L} = \f{2\pi^2 L^3}{\b^3}\left(-c_0 + 6c_1 \f{\b^2}{L^2}+ \cdots\right),
\ee
but now we have access to another quantization channel, from which we can extract the vacuum energy and write
\be
-\log Z\left(\f{3\pi L}{2}\right)_{S^1_\b \times S^2_{L/\sqrt{3}}} = \f{2\pi^2 L^3 }{\b^3}\left(-\evac+ 6\tilde{c}_1 \f{\b^2}{L^2}+\cdots\right) - \log \sum e^{- \f{3\pi L}{2} \Delta},
\ee
where $\D = E- E_{\text{vac}} \geq 0$. We can argue using the effective action that the sum is subleading. To see this, we consider a slightly more general partition function at high temperature $\b \rightarrow 0$, 
\be\label{quant1}
-\log Z(\b)_{S^1_{L_1} \times S^2_{L_2}}= \f{4\pi L_2^2 L_1}{\b^3}\left(-\evac + g(\b/L_2)\right) ,
\ee
where we captured all the perturbative sub-extensive corrections in $g(\b/L_2)$. Notice there are no $\b/L_1$ corrections due to the  effective action, as there is no curvature associated to the circle of size $L_1$. We also quantize along $L_1$ to find
\be\label{quant2}
-\log Z(L_1)_{S^1_{\b} \times S^2_{L_2}}  = \f{4\pi L_2^2 L_1 }{\b^3}\left(-\evac+ h(\b/L_2)\right) - \log \sum e^{- L_1 \D}\,.
\ee
In this more general situation, we can take $L_1 \rightarrow \infty$ to suppress the sum over excited states, in which case $g(\b/L_2) = h(\b/L_2)$. But this equality is independent of $L_1$, so is true even for $L_1 \sim L_2$, as we have in our problem above. This is similar to how, at leading order in $\b \rightarrow 0$, we project to the vacuum for a partition function $Z(L)_{S^1_\b \times \mathcal{M}}$, even though we don't take $L \rightarrow \infty$ (see Appendix A of \cite{Shaghoulian:2015kta} for details). But now using our effective action we see this is true for the first perturbative correction as well. 

Another way to see this is to take $L_1$ derivatives of both quantizations \eqref{quant1}-\eqref{quant2}, which shows $\partial_{L_1}\log Z(\b)_{S^1_{L_1} \times S^2_{L_2}}$ is $L_1$-independent and $\partial_{L_1}\log Z(L_1)_{S^1_{\b} \times S^2_{L_2}}$ is $L_1$-independent up to the piece $\partial_{L_1} \log \sum e^{-L_1 \D} = -\f{\sum \D e^{-L_1 \D}}{\sum e^{-L_1 \D}}$, which must therefore vanish. But the vanishing of this expression means $\log \sum e^{-L_1 \D}$ is $L_1$-independent, i.e. it vanishes.    Altogether we have that the series of corrections to the high-temperature free energy $\log Z(\b)_{S^1_{L_1} \times S^{2}_{L_2}}$ is equal to the series of corrections to the vacuum energy on $S^1_{\b} \times S^{2}_{L_2}$. 

Returning to the problem of interest, we find $\tilde{c}_1 = c_1$, i.e. the first correction to the vacuum energy on $S^1_\b \times S^2_{L/\sqrt{3}}$ exactly equals the first subextensive correction to the high-temperature free energy on $S^3_L$:
\be\label{secondbox}
\hspace{-6mm}\boxed{\text{to first subextensive order}:\quad -\log Z(\b)_{S^3_L} \approx \f{2\pi^2 L^3}{\b^3}\left(-c_0 + 6c_1 \f{\b^2}{L^2}\right) \approx \f{3\pi L}{2} \,E_{\text{vac}, S^1_\b \times S^2_{L/\sqrt{3}}}.}
\ee
All of this has generalizations to arbitrary dimension $d \geq 4$. For that we use
\begin{align}
\text{Vol}(S^{d-1}_L) &= \f{2\pi^{d/2}}{\Gamma[d/2]}\,L^{d-1}\,,\qquad &&\text{Ricci}(S^{d-1}_L) = \f{(d-1)(d-2)}{L^2}\,,\\
\text{Vol}(S_{L_1}^1 \times S_{L_2}^{d-2}) &= \f{2\pi^{(d-1)/2}}{\Gamma[(d-1)/2]}\,L_1L_2^{d-2}\,,\qquad &&\text{Ricci}(S_{L_1}^1 \times S^{d-2}_{L_2}) = \f{(d-2)(d-3)}{L_2^2}\,.
\end{align}
Equating and solving gives
\be
L_1 = \f{\left(\f{d-3}{d-1}\right)^{1-d/2}\sqrt{\pi} \,\Gamma\left[\f{d-1}{2}\right]}{\Gamma\left[\f d 2 \right]} \, L\,,\qquad L_2 = \sqrt{\f{d-3}{d-1}} \, L\,.
\ee
With these identifications we have
\be
\text{to first subextensive order}:\quad -\log Z(\b)_{S^{d-1}_L} \approx  L_1\,E_{\text{vac}, S^1_\b \times S^{d-2}_{L_2}}\,.
\ee
\subsection{Checks on modular invariance at subextensive order}
Without the skills to perform experiments to check our arguments, it is sometimes fun to perform some numerics. So let's do this. We will begin with a conformally coupled scalar field:
\be
I =  \int d^d x \, \sqrt{g}\left(\f 1 2\p^\m \phi \p_\m \phi + \f{d-2}{8(d-1)} \, R \phi^2\right).
\ee
It is relatively straightforward to check the first equality in \eqref{firstbox}, $ Z(\b)_{S^3_L} = Z(\b)_{S^1_{\f{3\pi L}{2}}\times S^2_{L/\sqrt{3}}}$ to first subextensive order. We will set $L = 1$. To calculate the thermal partition function on some spatial manifold $\mathcal{M}^{d-1}$, we need the eigenvalues of the Laplacian $\D = -\nabla^2_{\mathcal{M}^{d-1}} + \f{d-2}{4(d-1)} R$. This free theory has a zero mode that survives the Kaluza-Klein reduction and therefore contributes to the effective action, but this feature will not play a role in our analysis since the contribution appears at a subleading order in $\b$ than we are interested in. 

\begin{figure}
\centering
\includegraphics[scale=0.116]{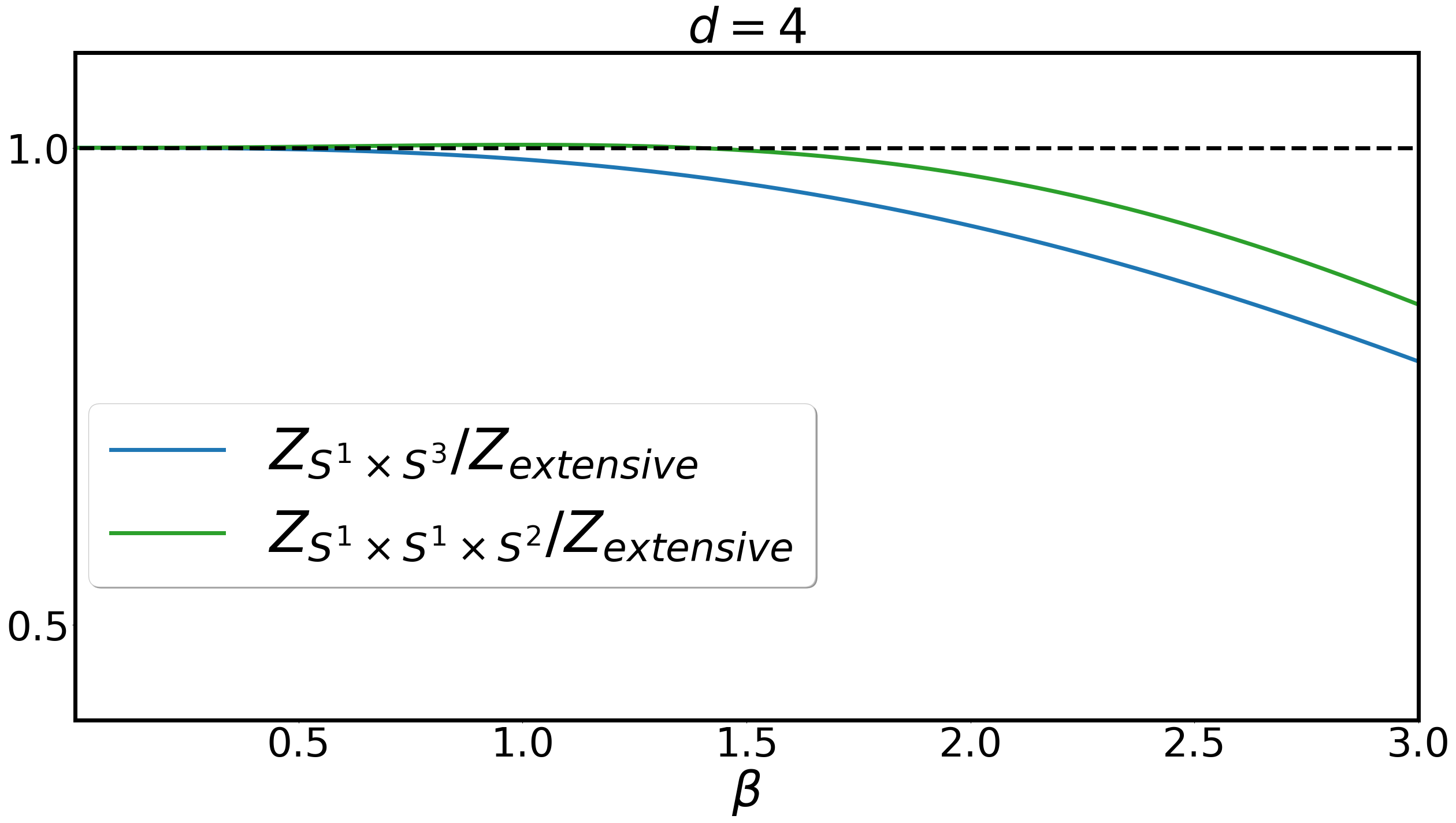}\qquad \quad  \includegraphics[scale=0.116]{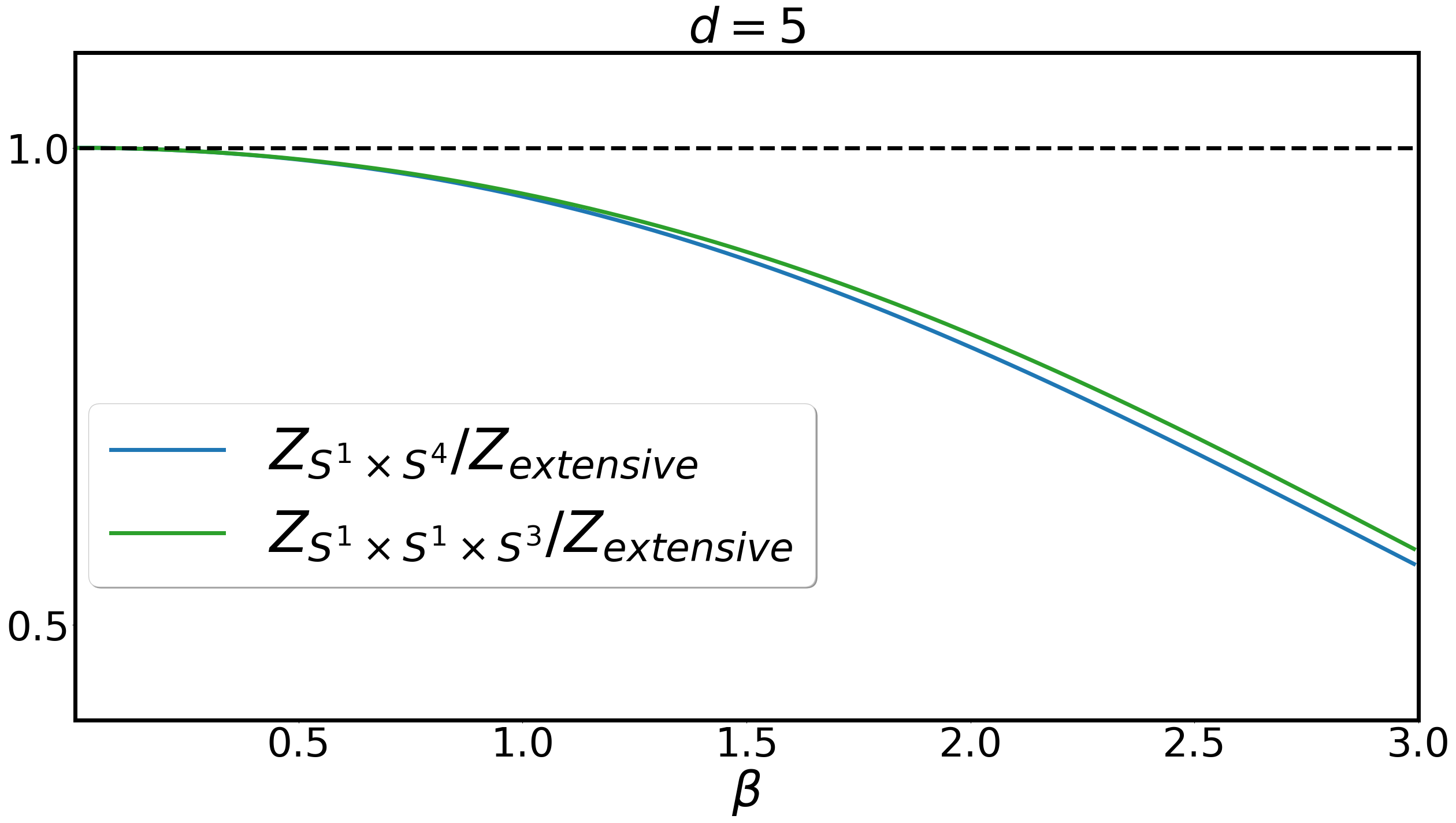}\qquad \includegraphics[scale=0.116]{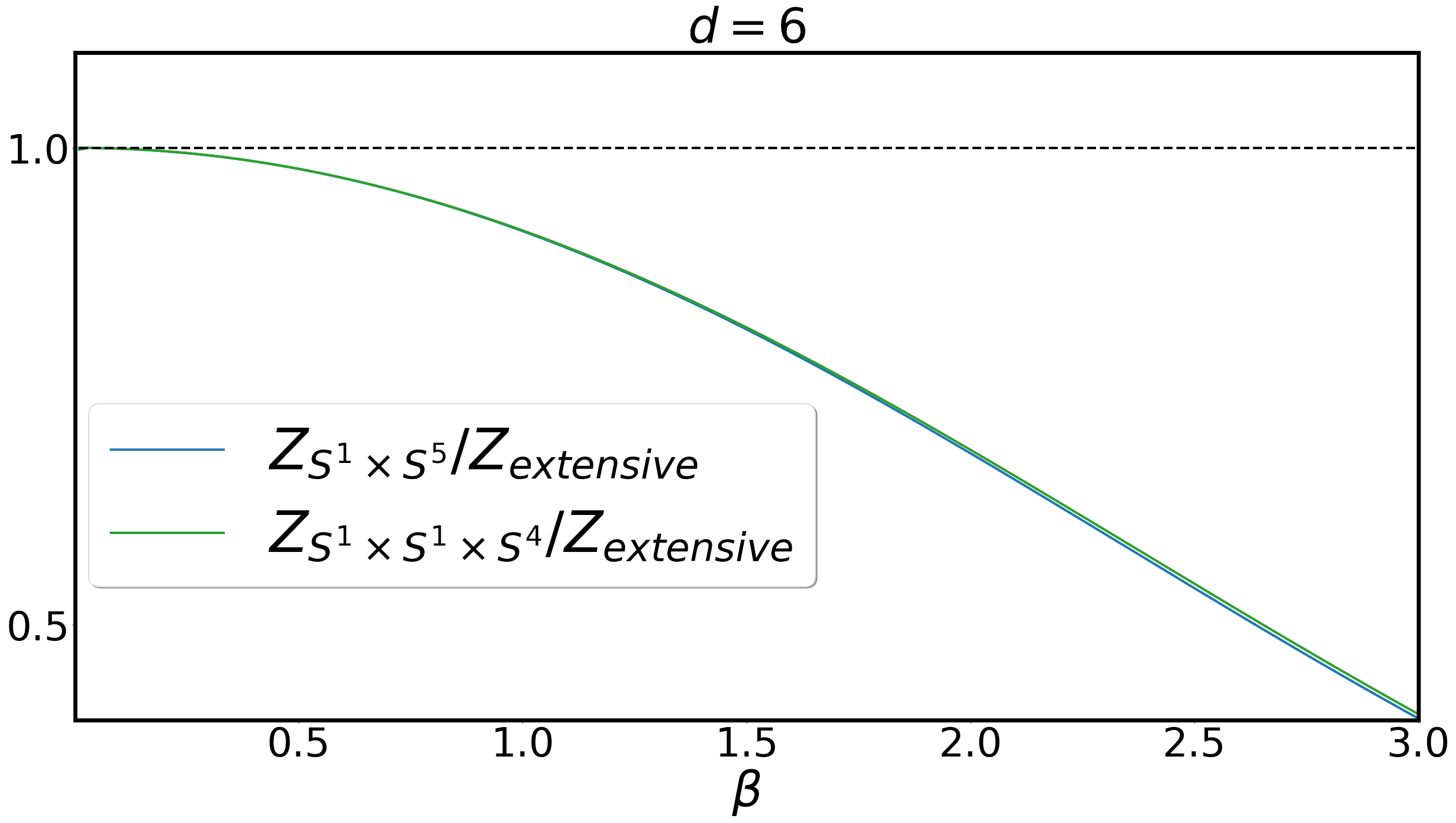} 
\caption{These plots represent a numerical calculation of the partition function for a conformally coupled scalar field in $d=4,\,5,\,6$ on a unit-radius $S^{d-1}$ and $S^1_{\f{3\pi}{2}} \times S_{1/\sqrt{3}}^{d-2}$. For $d=5,\,6$ the agreement between the two beyond the leading extensive term signals the agreement of the coefficient $c_1$. For comparison we also plot the $d=4$ case where there is no agreement beyond the leading extensive term, which is expected since $c_1 = 0$ in this case.}\label{numericalplots}
\end{figure}

We start with $S^3$, for which $R = (d-2)(d-1)$ and so the eigenvalues of the Laplacian $\D = -\nabla^2_{S^{d-1}} + (d-2)^2/4$ and their degeneracies are given as follows, see e.g. \cite{Giombi:2014yra}:
\be
\l_n = \w_n^2 = \left[n + \f 1 2 (d-2)\right]^2\,,\qquad d_n = (2n+d-2) \f{(n+d-3)!}{(d-2)!n!}
\ee
This gives
\be\label{s3i}
-\log Z = \sum_{n=0}^\infty d_n \log \left(1-e^{-\b \l_n}\right)\,.
\ee
On $S^1_{\f{3\pi}{2}}\times S^2_{1/\sqrt{3}}$, we have $R = 3(d-3)(d-2)$ and so the eigenvalues of the Laplacian $\D = -\p^2 -\nabla^2_{S^{d-2}} + 3(d-2)^2(d-3)/(4(d-1))$ and their degeneracies are given by 
\be
\hspace{-5mm}\l_{m,n} = \w_{m,n}^2 = \left(\f{4m}{3}\right)^2 + \left[3n(n+d-3) + \f{3(d-2)^2(d-3)}{4(d-1)}\right]^2,\quad d_n = (2n+d-3)\f{(n+d-4)!}{(d-3)!n!}\,.
\ee
This gives
\be\label{s1s2i}
-\log Z = \sum_{n=0}^\infty \sum_{m=-\infty}^\infty d_n \log\left(1- e^{-\b \l_{n,m}}\right).
\ee
We can now evaluate each of these numerically for small $\b$ by truncating the sums at some large number, and comparing them to the leading extensive piece of the conformally coupled scalar, known to be $-2\zeta(d)/\b^{d-1}$, where $\zeta$ is the Riemann zeta function. We do this for $d \geq 4$. As discussed in the previous section, it is already known that the leading pieces of \eqref{s3i} and \eqref{s1s2i} agree with $-2\zeta(d)/\b^{d-1}$. However, by evaluating numerically, one can check that \eqref{s3i} and \eqref{s1s2i} begin disagreeing with $-2\zeta(d)/\b^{d-1}$ once $\b$ becomes big enough, yet still agree with one another. This additional agreement is precisely that of the first subextensive term, and the eventual disagreement is when the second and higher order subextensive terms kick in. This additional agreement is not observed for $d = 4$, which is consistent with the fact that $c_1 = 0$ for the conformally coupled scalar in $d=4$. This means that the first correction beyond extensive order will be at linear order in $\beta$, which is the quadratic-in-curvature terms in the effective action -- there is no argument that these terms should agree, and indeed they do not. The numerical checks for $d=4, 5, 6$, are presented in Figure \ref{numericalplots}. 

To check the subextensive term of the Casimir energy, and therefore \eqref{secondbox}, would require regularizing the mode sum for the Casimir energy for $S^1 \times S^{d-2}$, which we have not done. Instead, we can check \eqref{secondbox} in holographic theories at strong coupling using the bulk description. The left-hand-side of \eqref{secondbox} is straightforward to calculate: this is simply the free energy of the spherical AdS-Schwarzschild black hole, expanded around high temperature. The right-hand-side, however, requires a vacuum geometry of the theory on $S^1 \times S^2$. The energy of this geometry would then give the Casimir energy of the theory on $S^1 \times S^2$. In fact, such a geometry was constructed numerically and proposed to give the vacuum state in \cite{Copsey:2006br} For the AdS$_5$ solution, the metric ansatz for a spacetime with $SO(3) \times SO(2)$ symmetry is 
\be
ds^2 = -e^{\g(r)} dt^2 + \a(r) d\chi^2 + \f{dr^2}{\a(r)\eta(r)} + r^2 d\Omega_2^2. 
\ee
The Einstein equations can be reduced to a single, third-order, nonlinear ordinary differential equation for $\a(r)$. This can be solved numerically, with the constants chosen such that the $\chi$ circle pinches off in the interior. This circle is interpreted as the spatial $S^1$ in the boundary theory. Using a different numerical algorithm, we have checked and used the values for the Casimir energy given in Table 1 of \cite{Copsey:2006br}. We can now compare this Casimir energy to the exact free energy, and to the purely extensive piece of the free energy, at some small inverse temperature $\b$. All three agree at sufficiently small $\b$, but as $\b$ is increased the Casimir energy and free energy continue to agree with one another but disagree with the purely extensive piece of the free energy. This is again the leading subextensive piece kicking in, and it is precisely telling us that this piece agrees between the Casimir energy on $S^1 \times S^2$ and the free energy on $S^3$. At even larger $\b$ they disagree with one another, as we reach the second and higher order subextensive pieces. 

These vacuum geometries were extended to higher dimensions in \cite{Harlow:2018tng}, and one can further check numerically that the subextensive piece of the Casimir energy of these geometries agrees with the subextensive piece of the free energy of the spherical AdS-Schwarzschild black hole, although we have not performed this check. 

\subsection{Evidence for $c_1 \geq 0$}\label{conjsec}
As stated previously, we conjecture that $c_1 \geq 0$. Let's now consider the evidence for this conjecture. We begin by considering holographic CFTs in $d$ dimensions, for which the computation of the free energy on $S^1_\b \times S^{d-1}_L$ reduces to the calculation of the action of an AdS-Schwarzschild black hole:
\be
ds^2 = -\left(1-\f{\m}{r^{d-2}}+r^2\right) dt^2 + \f{dr^2}{1-\f{\m}{r^{d-2}} + r^2} + r^2 d\Omega_{d-1}^2\,.
\ee
The temperature is given by
\be\label{brp}
\b = \f{4\pi r_+}{d r_+^2 + d-2}
\ee
while the action is given by 
\be\label{logz}
-\log Z = I = -\f{\b \text{Vol}(S^{d-1})}{16\pi G} \left(r_+^d -  r_+^{d-2}+ \b\text{-independent}\right)
\ee
Inverting \eqref{brp} and plugging into \eqref{logz} gives
\be
I = \f{\text{Vol}(S^{d-1})}{\b^{d-1}}\left(-c_0 + c_1 (d-1)(d-2) \b^2+\dots\right)\,,
\ee
where we used $R[S^{d-1}] = (d-1)(d-2)$. The leading piece is extensive, $c_0 > 0$, but it is also true that $c_1 > 0$ for $d \geq 3$:
\be
c_1 = \f{4^{d-4}\pi^{d-3}}{(d-2) d^{d-2} G} > 0\,.
\ee

We can also consider free conformal field theories. In four dimensions, for free scalars, fermions, and Maxwell fields we have $c_1 \geq 0$, with $c_1 = 0$ in the case of scalars. In six dimensions, for free scalars, fermions, and tensor fields we have $c_1 > 0$ \cite{Kutasov:2000td}. For fermions in arbitrary $d>2$ we have  $c_1 > 0$. For conformally coupled scalars we have $c_1 > 0$ for $d > 4$, $c_1 = 0$ for $d=4$ as stated above, and $c_1 < 0$ for  $d=3$ (see e.g. \cite{Melia:2020pzd}). The 3d conformally coupled scalar is so far the only example we have that $c_1$ is negative. In this case, however, we need to be more careful. Free theories can have a gapless sector upon dimensional reduction, so their complete free energy is not necessarily captured by the effective action of only background terms. We define
\be
f_1 =  c_1 + f_1^{\text{gapless}} \,,
\ee
where $f_1$ will refer to the coefficient of the leading subextensive term in the free energy, while $c_1$ is as usual the coefficient of the Ricci scalar term in the thermal effective action. This nuance is only relevant in $d=3$, where the gapless sector contributes at $\b^0$, which is the same order as the leading subextensive correction controlled by $c_1$ in the thermal effective action. For the $d=3$ scalar, we have $f_1 = f_1^{\text{gapless}}$ and $c_1= 0$ (see Appendix C of \cite{Benjamin:2023qsc}). In all the other free theories we considered we have $f_1^{\text{gapless}} = 0$.

%The fact that $c_1 = 0$ for the 4d free scalar could have been troubling: the existence of a weakly coupled fixed point may have led to $c_1 < 0$ by choosing the sign of the deformation appropriately. The fact that no such weakly coupled fixed points exist in 4d may be interpreted as further evidence for the bound \cite{zohar}. 

The conformally coupled scalar is a particularly compelling example: as $d$ is decreased, $c_1$ decreases, \emph{saturating} at $c_1 = 0$ for $d = 4$, and remains at zero for $d=3$ instead of becoming negative!  This seems to reflect a fundamental limitation in $c_1$ becoming negative. It would be interesting to further test $c_1 \geq 0$ for interacting fixed points in 3d.  

Given that $c_1 = 0$ for the 3d and 4d free scalar, it is interesting to calculate $c_1$ for the Wilson-Fisher fixed point. This is because a perturbative calculation can be sufficient to extract its sign, so a large-$N$ analysis or the epsilon expansion can be used. As it turns out, $c_1$ has been calculated at leading order in $N$ in the Wilson-Fisher fixed point, and found to vanish \cite{diatlyk}! (See also \cite{Bobev:2023ggk} for calculations of $c_1$ in interacting theories beyond the large-$N$ limit.)

Altogether, our conjecture can be refined to be consistent with the existing evidence as follows:\\
\\
\emph{Conjecture: The coefficient $c_1$ of the leading subextensive correction to the effective action coming from the gapped sector satisfies $c_1 \geq 0$. } \\
\\
We point out that for arbitrary theories in $d \geq 4$ and fully gapped theories in $d=3$, $c_1$ captures the entire leading subextensive correction to the free energy, which scales as $\b^{3-d}$. We should also point out that having a thermal compactification is crucial: as a counterexample for non-thermal compactifications, see for example the $\mathbb{Z}_2$-twisted scalar \cite{Benjamin:2023qsc}.

One way to try to prove that $c_1 \geq 0$ in $d \geq 4$ is by using the fact that $c_1$ appears in the Casimir energy. This has precedent, as the subextensive piece of the Casimir energy has been shown to have a sign constraint in \cite{Belin:2016yll}. In that paper, the subextensive piece was for a Casimir energy on the flat torus, so it referred to nonperturbative corrections to the thermal effective action (since all curvature terms in the effective action vanish), and the sign constraint gave the opposite sign, but nevertheless the fact that subextensive terms can be bounded is promising.

We recall that we can obtain sign constraints on the vacuum energy on $S^1 \times S^2$ as follows \cite{Levine:2022wos}. We extract the vacuum energy density by taking $\b \rightarrow \infty$:
\be\label{betaquant}
\b\rightarrow \infty: \quad - V_4^{-1} \log Z(\b)_{S^1_{L_1} \times S^2_{L_2}} \approx \f{ \b E_{\text{vac}}}{V_4} = \f{1}{L_2^4}g(L_1/L_2) = \f{1}{L_2^4}(-\tilde{\varepsilon}_{\text{vac}} + \tilde{g}(L_1/L_2))
\ee
where $V_4 = \text{Vol}(S^1_\b \times S^1_{L_1} \times S^2_{L_2}) = 4\pi L_2^2 L_1 \b$ and we have parameterized the vacuum energy in two different ways. We can quantize along the dual channel to get
\begin{align}
\b\rightarrow \infty: \quad-V_4^{-1} \log Z(L_1)_{S^1_{\b} \times S^2_{L_2}} &=- V_4^{-1} \log \left( e^{- L_1 E_{\text{vac}}} \sum e^{- L_1 \D}\right) \\
&=  \f{1}{L_2^4} g(\b/L_2) - \f{1}{4\pi L_2^2 L_1 \b} \log \sum e^{-L_1 \D}\,. \label{Lquant}
\end{align}
By equating this quantization with \eqref{betaquant} we can immediately get some constraints on $g(L_1/L_2)$. Extensivity of the the thermal entropy $-V_4^{-1}\log Z(L_1\rightarrow 0)_{S^1_\b \times S^2_{L_2}} \approx -\varepsilon_{\text{vac}}/L_1^4$, which tells us $g(L_1/L_2 \rightarrow 0) \approx -\evac\, \f{L_2^4}{L_1^4}$. The first derivative of \eqref{Lquant} with respect to $L_1$ is manifestly positive since $\D \geq 0$ by unitarity, while the second derivative is negative. This means $\p_{L_1} \tilde{g}(L_1/L_2) \geq 0$ and $\p^2_{L_1} \tilde{g}(L_1/L_2) \leq 0$. However, these sign constraints do not tell us about $c_1$. What we would like to do is parameterize the vacuum energy instead as 
\be\label{betaquant2}
\b\rightarrow \infty: \quad - V_4^{-1} \log Z(\b)_{S^1_{L_1} \times S^2_{L_2}} \approx \f{ \b E_{\text{vac}}}{V_4} =  \f{1}{L_1^4}(-\evac + f(L_1/L_2))
\ee
In this parameterization, a sign constraint on $f(L_1/L_2)$ (or sign constraint on its derivative) will immediately imply a sign constraint on $c_1$. Quantizing along the dual channel gives
\begin{align}
\b\rightarrow \infty: \quad -V_4^{-1} \log Z(L_1)_{S^1_{\b} \times S^2_{L_2}} &= -V_4^{-1} \log \left( e^{- L_1 E_{\text{vac}}} \sum e^{- L_1 \D}\right) \\
&=  \f{\evac}{\b^4} F(\b/L_2) -\f{1}{4\pi L_2^2 L_1 \b}\log \sum e^{-L_1 \D}\,.\label{Lquant2}
\end{align}
While the $L_1$ derivative  of \eqref{Lquant2} is positive, and therefore the $L_1$ derivative of \eqref{betaquant2} is positive, this does not lead to a sign constraint on $\p_{L_1} f(L_1/L_2)$. 

%In the case of the vacuum energy on $S^1_{L_1} \times T^2_{L_2}$, say where the torus $T^2$ has equal cycle sizes $L_2$, we are able to get a sign constraint precisely on the analogous function $f(L_1/L_2)$. In that case, however, the function $f(L_1/L_2)$ is not parameterizing EFT corrections, since those vanish on a flat manifold. It is instead parameterizing nonperturbative corrections. (Also importantly, the sign constraint that comes out of that approach is opposite the one we would want here.) 

In the rest of the paper we will restrict to theories without a gapless sector, $f_1 = c_1$.

\section{Angular velocity}\label{lorentztrick}
In this section we will discuss how the introduction of an angular potential affects thermodynamic quantities at leading extensive order. For a CFT on $\mathbb{T}^{d-1}$ this was studied in \cite{Shaghoulian:2015lcn}. We will work in the canonical ensemble and generalize those results to arbitrary angular velocities. For a CFT on $S^{d-1}$, when a fluid-dynamical approximation is valid, the result derived in \cite{Bhattacharyya:2007vs} is
\be\label{minwalla}
\log Z(\b, \vec{\Omega}) \equiv \log \Tr \exp\left[{-\b(H- \Omega_i J_i)}\right]\approx \f{c_0\text{Vol}(S^{d-1})}{\b^{d-1} \prod_i (1-\Omega_i^2)}
\ee
In other words, the leading-order answer without any angular velocity turned on is simply modified by the universal $\prod_i (1- \Omega_i^2)^{-1}$ factors. This was also recently derived from the effective field theory perspective\cite{Benjamin:2023qsc}, where the effective action is more involved since the metric has  a nontrivial Kaluza-Klein gauge field.

We will derive this  result for a generic conformal field theory using only Lorentz invariance, generalizing the technique for $\mathbb{T}^{d-1}$ in \cite{Shaghoulian:2015lcn}. The basic idea there was that the extensivity of the free energy at leading order in the temperature means that angular velocities can be incorporated simply through Lorentz transformations. Compact manifolds break Lorentz invariance globally, but for the extensive piece of the partition function the only effect this has is time dilation and length contraction, which change the size of thermal circle (time dilation) and the spatial manifold (length contraction). 

\subsection{CFT$_d$ on $T^{d-1}$}
\subsubsection{$T^{d-1}$: $d=2$}\label{secdeq2}
We will first consider a CFT$_2$ quantized on a circle of circumference $2\pi$. As is well-known, the two-dimensional Cardy formula gives the asymptotic density of states as 
\be
\rho(\D, J) \approx \exp\left(2\pi \sqrt{\f{c}{12}\left(\D + J - \f{c}{12}\right)}+2\pi \sqrt{\f{c}{12}\left(\D - J - \f{c}{12}\right)}\right)\,.
\ee
The density of states $\rho(\D, J)$ comes from an inverse Laplace transform of the high-temperature grand canonical partition
\be
Z(\b, \Omega) \equiv \Tr e^{-\beta (H - \Omega J)} \approx \exp\left(\frac{\pi^2}{3\beta(1-\Omega^2)}\right)\,,
\ee
where we take $\beta \rightarrow 0$ at fixed $\Omega$. 

A quick way to understand the appearance of the $(1-\Omega^2)^{-1}$ factor is through Lorentz invariance. As the high-temperature $\log Z$ is extensive, its density is equivalent to that of infinite flat spacetime. This means $\log Z$ at high temperature is Lorentz invariant up to a global effect of time dilation and length contraction. The angular velocity $\Omega$ is just the velocity, so we have the Lorentz gamma factor $\g = (1-\Omega^2)^{-1/2}$. We therefore introduce two gamma factors, one for time dilation and one for length contraction, to get
\be
\log Z(\b) \approx \f{\pi^2}{3\beta} \longrightarrow \f{\pi^2}{3\beta(1-\Omega^2)}\,.
\ee
To make this very explicit, let's re-introduce the length scale $L$ as the size of the spatial circle, and let's call $\beta_{nr}$ and $L_{nr}$ the size of the cycles for the non-rotating case, and $\beta$, $L$ the size of the cycles for the boosted case. Time dilation and length contraction give
\be
\b = \g \b_{nr} > \b_{nr}\,,\qquad L = L_{nr}/\g < L_{nr}\,,
\ee
which lets us write
\be
\log Z(\b_{nr}) = \f{\pi L_{nr}}{6\beta_{nr}} = \g^2 \f{\pi L}{6\b} = \f{\pi^2}{3\b (1-\Omega^2)}  = \log Z(\b, \Omega)\,,
\ee
where in the last equality we re-instated $L = 2\pi$.

\subsubsection{$T^{d-1}$: $d > 2$}
The case $d > 2$ has the additional subtlety that we can introduce multiple angular velocities $\Omega_i$, $i = 1, 2, \dots\,, d-1$. This corresponds to mutually orthogonal directions in which we can perform a boost. But in infinite flat space, we can simply add velocities and this amounts to a boost in \emph{some} direction.  In other words, we have a four-velocity $u^\a = (\g, \g\, \vec{v})$ with $\g = (1-\vec{v}\,^2)^{-1/2}$ and $v_i = R_i \Omega_i$ with $R_i = L_i/(2\pi)$ the radius of the $i$'th circle and $L_i$ its length. The overall volume of our spatial manifold will contract by one factor of $\g$ (i.e. the length in the single boost direction will contract), and the thermal cycle will dilate by one factor of $\g$. This means we have in the end
\be
\log Z(\b, \vec{\Omega}) \approx \f{c_0 \text{Vol}(\mathbb{T}^{d-1})}{\g^d \b^{d-1}} =\f{c_0 \text{Vol}(\mathbb{T}^{d-1})}{ \b^{d-1}(1- \sum_i R_i^2 \Omega_i^2)^{d/2}} \,.
\ee
The inverse Laplace transform of $Z(\b, \vec{\Omega})$ gives $\rho(E, \vec{J})$:
\be
\rho(E, \vec{J}) = \int \left(\prod_i d\Omega_i\right)d\b\, Z(\b, \vec{\Omega}) e^{\b (E - \Omega_i J_i)} .
\ee
This can be evaluated by saddle point, with the saddle-point values given implicitly as
\be
J_a = \f{c_0 \text{Vol}(\mathbb{T}^{d-1}) d R_a^2 \Omega_a}{\b^d (1- \sum_i R_i^2 \Omega_i^2)^{d/2+1}}\,,\qquad E = \f{c_0 \text{Vol}(\mathbb{T}^{d-1}) (d-1+\sum_i R_i^2 \Omega_i^2)}{\b^d (1- \sum_i R_i^2 \Omega_i^2)^{d/2+1}}\,.
\ee
Stated in terms of physical velocities $v_i = R_i \Omega_i$, we have
\be\label{torusd}
\log Z(\b, \vec{v}\,)  =\f{c_0 \text{Vol}(\mathbb{T}^{d-1})}{ \b^{d-1}(1- \vec{v}\,^2)^{d/2}} \,;\qquad J_a = \f{c_0 \text{Vol}(\mathbb{T}^{d-1}) d R_a v_a}{\b^d (1- \vec{v}\,^2)^{d/2+1}}\,,\quad E = \f{c_0 \text{Vol}(\mathbb{T}^{d-1})(d-1+ \vec{v}\,^2)}{\b^d (1- \vec{v}\,^2)^{d/2+1}}\,.
\ee
For a single angular momentum $J$ the result is given explicitly in \cite{Shaghoulian:2015lcn}.

\subsection{CFT$_d$ on $S^{d-1}$}\label{spherelorentz}
Now we use the same logic to derive \eqref{minwalla}. Importantly, we will not be making a fluid-dynamical approximation as in \cite{Bhattacharyya:2007vs}, so our result will apply to all CFTs, just as in \cite{Benjamin:2023qsc}. For $d=2$ this just corresponds to section \ref{secdeq2} above, so we begin with $d=3$. The complication compared to a flat torus is that different latitudes of a sphere rotating around the $z$-axis have different speeds, so we will need to integrate the Lorentz boost factor over the sphere. 
\subsubsection{$S^{d-1}$: $d=3$}
We can write the metric of our Euclidean manifold as 
\be
ds^2 = d\t^2 + d\q^2 + \sin^2 \q \,d\phi^2\,.
\ee
We can only have a single angular velocity in three dimensions, and it corresponds to a rotation in the $\phi$ direction: $\Omega = d\phi/dt$. The  physical velocity of a circle of latitude at angle $\theta$ is given by $v = \Omega \sin \q$. The high-temperature partition function with zero angular velocity is extensive, and we have to integrate up the Lorentz contraction factors $\g = (1-v^2)^{-1/2}$ on the circles of latitude to see what happens to the volume of the entire sphere:
\be
\log Z(\b_{nr}) \approx \f{4\pi c_0}{\b_{nr}^2} = c_0 \int_0^\pi d\q\int_0^{2\pi}d\phi \, \f{\sin \q}{\b_{nr}^2} \longrightarrow  c_0 \int_0^\pi d\q\int_0^{2\pi\g}\hspace{-3mm}d\phi \, \f{\g^2\sin \q}{\b^2}\,.
\ee
\vspace{-3mm}
\be
=  \int_0^\pi d\q \f{2\pi c_0 \sin\q}{(1-\Omega^2 \sin^2 \q)^{3/2}} = \f{4\pi c_0}{\b^2(1-\Omega^2)} = \log Z(\b, \Omega)\,.
\ee
\subsubsection{$S^{d-1}$: $d=4$}
In this case we can have two independent directions of rotation. Before treating that case let's turn on just a single angular velocity. We can write our Euclidean metric is 
\be
ds^2 = d\t^2 + d\q^2 + \sin^2 \q (d\psi^2 + \sin^2 \psi d\phi^2)\,
\ee
and consider an angular velocity $\Omega = d\phi/dt$ which gives a physical velocity $v = \Omega \sin \q \sin \psi$. We therefore transform as in the previous subsection
\be
\log Z(\b_{nr}) \approx c_0 \int_0^\pi d\q \int_0^\pi d\psi \int_0^{2\pi} \f{\sin^2 \q \sin \psi }{\b_{nr}^3} \longrightarrow c_0 \int_0^\pi d\q \int_0^\pi d\psi \int_0^{2\pi \g} d\phi \f{\g^3 \sin^2 \q \sin \psi}{\b^3}
\ee
\vspace{-3mm}
\be
 = \int_0^\pi d\q \int_0^\pi d\psi \f{2\pi c_0\sin^2 \q \sin \psi}{\b^3(1-\Omega^2 \sin^2 \q \sin^2\psi)^2} = \f{2\pi^2c_0}{\b^3(1-\Omega^2)}\,.
\ee
Now let's turn on our two independent directions of rotation. We can write our metric as 
\be
ds^2 = d\t^2 + d\q^2 + \sin^2 \q\, d\phi_1^2 + \cos^2 \q \,d\phi_2^2 \,,\qquad \q \in [0, \pi/2]\,,\quad \phi_i \sim \phi_i + 2\pi\,.
\ee
The angular velocities are $\Omega_i = d\phi_i/dt$, which correspond to physical velocities $v_1 = \Omega_1 \sin \q$, $v_2 = \Omega_2 \cos \q$ and $\g = (1-v_1^2 - v_2^2)^{-1/2}$. We can treat this case similarly to the $d>2$ case for a CFT on $\mathbb{T}^{d-1}$. In particular, there is some oblique direction between $\phi_1$ and $\phi_2$ along which we are boosting, and this will give us one $\g$ factor of length contraction in the volume. The thermal circle size $\b$ comes with its own factor of $\g$ as usual, and we have
\be\label{boost4d}
\log Z(\b_{nr}) \approx \f{2\pi^2 c_0}{\b_{nr}^3}\longrightarrow c_0 \int_0^{\pi/2} d\q\int_0^{2\pi}\int_0^{2\pi} \g\, d\phi_1 d\phi_2 \f{\g^3 \sin \q \cos \q}{\b^3} = \f{2\pi^2 c_0}{\b^3(1-\Omega_1^2)(1-\Omega_2^2)}\,.
\ee
\subsubsection{$S^{d-1}$: general $d$}
The general rule is now clear. We can turn on several independent angular velocities $\Omega_i$, which will correspond to some physical velocities $v_i$ and $\g = (1-\sum v_i^2)^{-1/2}$. Now we simply do the $(d-1)$-sphere volume integral with an additional $\g^d$ inserted in the integrand, where $(d-1)$ factors come from time dilation $\b^{1-d} \rightarrow \g^{d-1}\b^{1-d}$ and one factor comes from length contraction. Doing the integral then results in 
\be
\log Z(\b, \Omega_i) \approx \f{c_0\text{Vol}(S^{d-1})}{\b^{d-1} \prod_i (1-\Omega_i^2)}\,.
\ee
Mathematically, the integrals needed for this approach turn out to be the same as in the EFT approach \cite{Benjamin:2023qsc}, which we review in Appendix \ref{app}.

%At first subleading order we have EFT terms $R[S^2]$ and $F^2 = h^{\q\q} h^{\phi\phi} (\p_\theta A_\phi)^2 = 2$, so the nontrivial fibration enters in the lens space quantization channel, whereas in the $S^1 \times S^3$ channel we have just an $R[S^3]$ term in the EFT. We can try turning on an angular velocity in the $S^1 \times S^3$ channel, but then we have to modify $\beta$ to match the universal $(1-\Omega^2)^{-1}$ factor which appears at leading order in the $S^1 \times S^3$ quantization channel. We have one constraint on $\b$, $\Omega$ to match the leading order free energy, but now we have two terms at first subleading order in either quantization channel, and we do not have enough freedom to independently match both. 

\section{Further applications of thermal effective action}
\subsection{Modular invariance on $S^1 \times S^3/\mathbb{Z}_p$}
In \cite{Shaghoulian:2016gol}, the following modular  relation was proposed:
\be\label{conjshag}
\lim_{p \rightarrow 0}: \qquad \log Z(S^1_{2\pi/p} \times S^3) \approx \log Z(S^1_{2\pi} \times S^3/\mathbb{Z}_p).
\ee
The equality was between the extensive $1/p^3$ type pieces. Stronger relations were proposed in that paper but we will focus on this one, and actually derive it from the effective theory. The surprising element is that the nontrivial fibration of the $S^1$ in the channel with $S^3/\mathbb{Z}_p$ washes out at order $1/p^3$.

We write the metric for a unit-radius lens space as  
\be
ds^2 = \f 1 4\left[d\q^2 + \sin^2 \q \, d\phi^2 + (d\psi + \cos \q \, d\phi)^2\right],\qquad \psi \sim \psi + 4\pi/p\,.
\ee
We can do a constant Weyl transformation to ignore the overall factor of $1/4$, and we write our metric as 
\be
ds^2 = d\psi^2 + 2 \cos^2 \q \, d\psi d\phi + d\q^2 + d\phi^2\,.
\ee
Comparing to the Kaluza-Klein form \eqref{kkmetric} we are able to read off
\be
A_\phi = \cos^2 \q\,,\qquad h_{\q\q} = 1\,, \quad h_{\phi\phi} = 1- A_\phi^2 = \sin^2 \q\,.
\ee
In particular this means the 2d metric is just that of an $S^2$, so the fibration washes out at leading order. This proves the conjecture \eqref{conjshag}. There is a supersymmetric generalization of this conjecture relating a high-temperature supersymmetric index \cite{DiPietro:2014bca} to a supersymmetric Casimir energy on the sphere  \cite{Shaghoulian:2016gol}. It would be interesting to prove this by similar techniques. 
\subsection{Higher order curvature terms}
It is natural to ask whether our trick of matching the linear-in-$R$ term by exchanging $S^{d-1}$ with $S^1 \times S^{d-2}$ can be extended to further terms in the effective action. This is difficult due to the proliferation of terms at higher orders in the expansion. For example, for the case of CFT$_4$, at $O(\b^2)$ we will have both $R^2$ and $R_{\m\n} R^{\m\n}$. While the $R^2$ term will be matched due to the $R$ term being matched,  we do not have any freedom left in our spatial $S^1 \times S^2$ to match the $R_{\m\n} R^{\m\n}$ term. 

To introduce more freedom, we can consider the theory at finite temperature on a squashed sphere (aka Berger sphere) $S^3_\n$ instead of an ordinary sphere. Then we can match the Ricci scalar and $R_{\m\n}R^{\m\n}$ with $S^1_{L_1} \times S^2_{L_2}$ by tuning $\nu$:
\begin{align}
S^3_\nu:& \qquad R^2 = \f{4(\nu^2 - 4)^2}{L^4}\,,\qquad R^{\m\n}R_{\m\n} = \f{4(8-8\nu^2 + 3\nu^4)}{L^4}\\
S^1_{L_1} \times S^2_{L_2}:&\qquad R^2 = \f{4}{L_2^4}\,,\qquad \qquad\quad R^{\m\n}R_{\m\n} = \f{2}{L_2^4}\,.
\end{align}
Solving these gives
\be
L_2 = \f{L}{\sqrt{\n^2-4}}\,,\qquad \nu =  2\sqrt{2/5}
\ee
We can also scale $L_1$ to match the volume. Altogether this means the terms at $O(\b^{-2})$, $O(\b^0)$ and $O(\b^2)$ in the free energy for a CFT$_4$ on $S^3_\n$ with $\n = 2 \sqrt{2/5}$ can be matched with the terms of the same order in the free energy on $S^1 \times S^2$.

Another route to introducing more freedom is to go to higher dimensions, so that we can replace $S^1 \times S^{d-1}$ with $S^1 \times S^2 \times \cdots$ or e.g. $S^1 \times S^4$ with $S^1 \times S^1 \times S^3_\nu$. Restricting to the case $d \leq 6$ does not give us much mileage, although this may be an effective route for studying the $d \rightarrow \infty$ limit.
\subsection{$U(1)$ charge}
We can consider the following grand canonical ensemble:
\be
Z(\b, \Omega, \Phi) = \Tr e^{-\b(H - \Omega J - \Phi Q)}
\ee
where $Q$ is a conserved $U(1)$ charge and $\Phi$ is the chemical potential, which has dimensions of an inverse length (since $Q$ is dimensionless as a conserved charge).  Our EFT therefore has to be generalized:
\be
-\log Z(\b, \Omega, \Phi) = \int d^{d-1} \Sigma\left(-c_0\f{1}{\b^{d-1}} + c_1\f{R}{\b^{d-3}}  + c_2 \f{F^2}{\b^{d-3}} + c_3\f{\Phi^2}{\b^{d-3}}+ \cdots \right) +\text{nonpert.}
\ee
Dimensional analysis would let us write a term like $\f{\Phi}{\b^{d-2}}$, but we eliminate all odd powers of $\Phi$ due to the existence of antiparticles. In other words, for every state of charge $Q$ there is a state of charge $-Q$, so the partition function obeys $Z(\b, \Phi) = Z(\b, -\Phi)$. At first subextensive order in our EFT we now have three independent terms. %We won't bother going to higher order since we can't say anything interesting about those terms even for $\Phi = 0$. 
The $\Phi^2$ term can be matched between the $S^{d-1}$ and $S^1 \times S^{d-2}$ frames simply by picking $\Phi$ on $S^1 \times S^{d-2}$ to be the same as on $S^{d-1}$. 

We can compute these coefficients for holographic theories (see also \cite{Benjamin:2023qsc}). We have
\begin{align}
c_0 &=\f{1}{4Gd} \left(\f{4\pi}{d}\right)^{d-1}\,,\\
c_1 & = \f{1}{4 G d(d-2)}\left(\f{4\pi}{d}\right)^{d-3}\,,\\
c_2 & =  -\f{1}{32 G(d-2)} \left(\f{4\pi}{d}\right)^{d-3}\,,\\
c_3 & = -\f{d-2}{2 G d}\left(\f{4\pi}{d}\right)^{d-3}\,.
\end{align}

\subsection{Correlation functions}
Similar to sourcing conserved currents, we can source a generic operator. For simplicity let's stick to a scalar operator $\mathcal{O}$ with dimension $\D \leq d$, which has source $\tilde{J}$ with dimension $d-\D$. This is a relevant (or marginal in the case of $\D = d$) operator and appears in the path integral as 
\be
Z = \int [D\psi] e^{-S + \int d^d x \tilde{J}(x) O(x)}\,.
\ee
The thermal effective action can now be written in terms of the background terms $R_{\m\n\r\s}$ and $J$:
\be
-\log Z = \int d^{d-1} x \sqrt{g} \left(-c_0 \f{1}{\b^{d-1}} + c_1 \f{R}{\b^{d-3}} + \cdots + \f{J(x)}{\b^{\D}} \left(a_0 + a_1 \f{R}{\b^{\D-2}} + \cdots\right) % + \f{b_1 J(x)^2}{\b^{2\D-d+1}} +\f{b_2 \Box J}{\b^{\D-2}}
+ \cdots \right )
\ee
Notice that due to the dimensional reduction $[J] = d-\D-1$. The 1-pt function is given by 
\be
\f{\d \log Z}{\d J} = \langle O \rangle = \f{1}{\b^\D}\left(a_0 + a_1 \f{R}{\b^{\D-2}}+ \cdots \right)
\ee
So we see that the one-point function steps down by powers of $2$ in a small-$\b$ expansion. By an inverse Laplace transform, this is related to an average over the light-heavy-heavy OPE coefficients. 

Higher-point functions are more difficult to access with the technique above.\footnote{Thanks to Zohar Komargodski for discussion about this case.} This is because the effective theory only captures correlators of operators uniform in the time direction, like the Kaluza-Klein zero mode:
\be
Z[J] = Z[0] \sum_{n=0}^\infty \f{1}{n!} \int d^d x_1 \cdots \int d^d x_n \tilde{J}(x_1)\cdots \tilde{J}(x_n) \langle \mathcal{O}(x_1)\cdots \mathcal{O}(x_n)\rangle \,.
\ee
We are interested in the thermal vacuum on $S^{d-1}$. We can write $\mathcal{O}(\t, x) = \sum e^{2\pi i n \t/\b}\mathcal{O}_n (x)$ and do the $\t$ integrals. This projects to the $n=0$ mode and therefore picks up the correlators of the KK zero modes $\mathcal{O}_0(x)$. This is fine for the 1-point function $\langle 0| \mathcal{O}(\t,x) |0\rangle = \langle 0|\mathcal{O}|0\rangle $, since it is independent of time (and space), and therefore must equal the 1-point function of the KK zero mode $\mathcal{O}_0(x)$. 

%Equation (100) of https://arxiv.org/pdf/2211.05144.pdf seems to step down by two powers of $T$. 

\subsection{Holographic theories}
Holographic theories are a particularly interesting application of the thermal effective field theory techniques, as  the effective action has an extended range of validity \cite{Horowitz:2017ifu}. While for a generic theory it is only expected to hold asymptotically at small $\b$, for holographic theories the effective action is accurate up to $\b \sim O(1)$, with the precise numerical factor being determined by the Hawking-Page phase transition temperature in the bulk. This transition indicates a nonperturbative correction to the effective action which is suppressed by $N$. 

In the case of CFT$_3$, the thermal effective action is particularly simple, since a 2d geometry is characterized completely by the Ricci scalar. This means the effective action is written in terms of powers and derivatives of the Ricci scalar. The part of the effective action without any derivatives is given by
\be
-\log Z(\b) \supset \int d^2 x \sqrt{h} \left(-c_0\f{1}{\b^2} + c_1 R + \sum_{i=2}^\infty a_i R^i \b^{2(i-1)} \right)\,.
\ee
We can solve for the $a_i$ by expanding the free energy of the Schwarzschild-AdS black hole. We find
\be
c_0 = \f{4\pi^2}{27}\,,\qquad c_1 = \f{1}{12}\,,\qquad a_{n \geq 2} = -\f{3^{n-2} \pi^{3/2-2n} \G[n-3/2]}{2^{3n+1}  n!} \,.
\ee
These Wilson coefficients will be part of the answer for the free energy density of the CFT on any curved 2d manifold, and therefore the free energy density of the corresponding black hole in the bulk. But this is not very useful, since starting at order $\b^4$ we can have derivative terms, for example $R \Box R$, which vanish for $S^2$ but are activated for a general curved manifold. This means that we cannot predict the value of the free energy density at this and higher orders simply from the effective action on $S^2$. One way to try to make progress is to solve the equations for a black hole with a horizon that is slightly aspherical, by adding a generic, linearized perturbation. Computing the free energy of this solution will give information about the coefficients of all the derivative terms in the effective action. These coefficients can then be used for 2d geometries which are far from spherical, and for holographic theories the effective theory will hold for a wide range of $\b$. 

Going to higher dimensions complicates matters further: in $d = 4$ $R_{\m\n}$ is activated, and for $d\geq 5$ $R_{\m\n}$ and $R_{\m\n\rho\s}$ are activated. The results on a sphere geometry are not enough to disambiguate these from the Ricci scalar. 

%One way to make progress in the $d=3$ case is to perturb the horizon of the spherical black hole with a generic, linearized perturbation, and compute the free energy. 

\section{Conclusions}
In this paper we used thermal effective field theory to equate $c_1$ -- the coefficient of the leading subextensive correction to the free energy at inverse temperature $\b$ on $S^{d-1}$ -- to the coefficient of the leading subextensive correction to the Casimir energy on $S_\b^1 \times S^{d-2}$, for $d \geq 4$. This generalizes the relation at leading order in the free energy, where $c_0$ -- the coefficient of the extensive piece of the free energy -- is equated with the leading piece of the Casimir energy on $S^1_\b \times \mathcal{M}^{d-2}$ at small $\b$. The leading order relation deals with extensive formulas, in which case the manifold $\mathcal{M}^{d-2}$ is irrelevant as long as its volume is appropriately normalized. The subextensive correction, however, cares about the (integrated) Ricci scalar, so we have to restrict the class of manifolds we can consider; $S^{d-2}$ forms the simplest example, which is the one we studied in this paper. We checked this general argument against free theories and holographic theories, finding the expected agreement. 

We conjectured that $c_1 \geq 0$ in general dimension $d \geq 3$, although we were not able to prove this inequality using the techniques of \cite{Belin:2018oza}, which were used in the past to provide sign constraints for subextensive pieces of the Casimir energy. More traditional dispersive arguments may be useful here. A significant check of this constraint would be to calculate the leading subextensive correction to the Wilson-Fisher fixed point in $d = 3$. Since $c_1 = 0$ for the $d=3, 4$ conformally coupled scalar, perturbative methods like the epsilon expansion or the large $N$ expansion would suffice to probe a potential violation. 

Upon including angular velocity, we showed that the modification of the leading extensive piece of the free energy on $S^{d-1}$ can be derived simply by using Lorentz invariance, in particular by appropriately including time dilation and length contraction. In appendix \ref{app} we reviewed the subextensive corrections coming from the thermal effective action, as first done in the beautiful work of \cite{Benjamin:2023qsc}. Finally, we considered various extensions to the thermal effective action approach, for example proving a conjecture about modular invariance on $S^1 \times S^3/\mathbb{Z}_p$ from \cite{Shaghoulian:2016gol} and discussing the inclusion of additional chemical potentials and applications to holographic theories. 

It is interesting to imagine gauging the diffeomorphism symmetry, i.e. thinking of the thermal effective action as a theory of gravity. In that case our sign constraint $c_1 \geq 0$ maps to an Einstein-Hilbert term of the wrong sign. This means that the matter fields generate an Einstein-Hilbert term of the wrong sign, which may have implications for induced gravity scenarios. 

It would also be interesting to study sign constraints on the other Wilson coefficients. For the holographic theories we have $c_2 < 0$ and $c_3 < 0$. The coefficient $c_2$ was calculated in free theories and the 3d Ising model in \cite{Benjamin:2023qsc}, and was found to be negative in all those examples.  So it is natural to conjecture $c_2 \leq 0$ as another universal constraint.  %Furthermore, it is known that causality constraints can then be used to bound coefficients of various terms in this effective action, as in \cite{Adams:2006sv, Camanho:2014apa}.

\section*{Acknowledgments}
The authors would like to thank Nathan Benjamin, Jaeha Lee, Zohar Komargodski, Raghu Mahajan, Hirosi Ooguri, Sanjit Shashi, and David Simmons-Duffin for helpful conversations. The authors are supported in part by DOE grant DE-SC001010.

\appendix
\section{Effective field theory derivation of free energy}\label{app}
The results of section \ref{lorentztrick} can be obtained from thermal effective field theory. We will analyze $\mathbb{T}^{d-1}$  before turning to $S^{d-1}$, the case treated in \cite{Benjamin:2023qsc}. The effective action needs to be modified to include a gauge field $A_\m$ for the twist in the metric corresponding to an angular velocity (we will give more details on this parameterization below). The effective action is then written in terms of gauge-invariant combinations of the metric $g_{\m\n}$ and the gauge field $A_\m$:
\be
I = -\log Z = \int d^{d-1} \Sigma \left(-c_0 \f{1}{\b^{d-1}} + c_1 \f{R}{\b^{d-3}} + c_2 \f{F^2}{\b^{d-3}}+\cdots\right) + \text{non-pert.}
\ee
We will be overly explicit in the calculations below so that it is clear how they go. 

\subsection{CFT$_d$ on $T^{d-1}$}
Turning on angular velocities means we need to do the path integral of our CFT on the manifold
\be
ds^2 = d\t^2 + d\q^2 + d\phi_i^2\,,\qquad (\t, \phi_i) \sim (\t + \b, \phi_i + \Theta_i) \sim (\t, \phi_i + 2\pi R_i)\,,
\ee
where we have angular potentials $\Theta_i = i \b \Omega_i$. In the Hamiltonian formalism we are calculating
\be
Z(\b, \Theta) = \Tr \left[e^{-\beta H + i \Theta_i J_i}\right]\,.
\ee
To do the dimensional reduction we have to put the metric in Kaluza-Klein form:
\be\label{kkmetric}
ds^2 = d\t^2 + (h_{ij} + A_i A_j) dx^i dx^j + 2 A_i d\t dx^i\,,\qquad \t \sim \t + \b\,.
\ee
To achieve this, we define new angular coordinates
\be
\phi_i \longrightarrow \phi_i + \f{\Theta_i}{\b} \t \implies (\t, \phi_i) \sim (\t + \b, \phi_i + \Theta_i) \rightarrow (\t, \phi_i) \sim (\t + \b, \phi_i)\sim (\t, \phi_i+2\pi R_i)\,
\ee
which gives a metric
\be
ds^2 = d\t^2 + \left(d\phi_i + \f{\Theta_i}{\b} d\t\right)^2 = d\t^2 \left(1+\f{\Theta_i^2}{\b^2} \right)+ d\phi_i^2 + 2 \f{\Theta_i}{\b} \, d\t d\phi^i
\ee
\vspace{-3mm}
\be
\implies ds^2 \cong d\t^2 + \f{d\phi_i^2 + 2 \f{\Theta_i}{\b}  \, d\t d\phi^i}{1 + \f{\Theta_i^2}{\b^2}}
\ee
where in the last line we did a Weyl transformation. This isn't necessary but we find it convenient. Since we are working in a conformally flat geometry we will ignore the possibility of Wess-Zumino terms for the Weyl anomaly.  Comparing to \eqref{kkmetric} lets us  identify
\be
A_i =\f{\Theta_i/\b}{1+\Theta_k^2/\b^2}\,,\qquad h_{ij} = \begin{cases}\f{1}{1+\Theta_k^2/\b^2} - A_a A_a =  \f{1+\sum_{n\neq a} \Theta_n^2/\b^2 }{\left(1+\Theta_k^2/\b^2\right)^2}\,,\qquad i = j = a\\ -A_i A_j \hspace{16mm}  = - \f{\Theta_i \Theta_j/\b^2}{\left(1+\Theta_k^2/\b^2\right)^2}\,,\qquad\hspace{1mm} i \neq j\end{cases}.
\ee
Notice there is no sum over the repeated $a$ index. The thermal effective action only has the leading cosmological constant term, since $R = 0$ and $F_{\m\n}$ = 0 (as the gauge field $A_\m$ is a constant). Substituting $\Theta = i \b R \Omega$ for radius $R$ to write the answer in terms of the angular velocity gives us 
\be
\hspace{-4mm} \log Z(\b, \Omega_i) = c_0 \int d^{d-1} x \f{\sqrt{h} }{\b^{d-1}} = \f{c_0}{\b^{d-1} \left(1+ \Theta_i^2/\b^2\right)^{d/2}} \int  d^{d-1} x = \f{c_0 \text{Vol}(T^{d-1})}{\b^{d-1}\left(1-R_i^2\Omega_i^2\right)^{d/2}}\,.
\ee
This reproduces \eqref{torusd}. 

\subsection{$CFT_d$ on $S^{d-1}$}
In this section we apply the same EFT technique for the theory quantized on a sphere, at finite temperature and finite angular potential. These results are already in \cite{Benjamin:2023qsc}, but we rederive them here so we can compare to the results from Section \ref{spherelorentz}. 

\subsection*{{\bf $d=3$}} 
Turning on an angular velocity means we want to do the path integral of our CFT on the manifold
\be
ds^2 = d\t^2 + d\q^2 + \sin^2 \q \, d\phi^2\,,\qquad (\t, \phi) \sim (\t + \b, \phi + \Theta) \sim (\t, \phi + 2\pi)\,,
\ee
where we have an angular potential $\Theta = i \b \Omega$. We want to calculate
\be
Z(\b, \Theta) = \Tr \left[e^{-\beta H + i \Theta J}\right]\,.
\ee
To put the metric in Kaluza-Klein form \eqref{kkmetric}, we define a new angular coordinate
\be
\phi \longrightarrow \phi + \f{\Theta}{\b} \t \implies (\t, \phi) \sim (\t + \b, \phi + \Theta) \rightarrow (\t, \phi) \sim (\t + \b, \phi)\,
\ee
which gives 
\be
ds^2 = d\t^2 + d\q^2 + \sin^2 \q \left(d\phi + \f{\Theta}{\b} d\t\right)^2 = d\t^2 \left(1+\f{\Theta^2}{\b^2} \sin^2 \q\right) + d\q^2 + \sin^2 \q \, d\phi^2 + 2 \f{\Theta}{\b} \sin^2 \q\, d\t d\phi
\ee
\vspace{-3mm}
\be
\implies ds^2 \cong d\t^2 + \f{d\q^2 + \sin^2\q \, d\phi^2 + 2 \f{\Theta}{\b} \sin^2 \q \, d\t d\phi}{1 + \f{\Theta^2}{\b^2} \sin^2 \q}
\ee
where in the last line we did a Weyl transformation. Writing \eqref{kkmetric} as
\be
ds^2 = d\t^2 + h_{\q\q} d\q^2 + \left(h_{\phi\phi} + A_{\phi}^2\right) d\phi^2 + 2 A_{\phi} d\t d\phi
\ee
we can identify
\be
A_i = (A_\q, A_{\phi}) = \left(0, \f{\f{\Theta}{\b} \sin^2 \q}{1 + \f{\Theta^2}{\b^2} \sin^2 \q}\right)\,,\qquad h_{ij} = \begin{pmatrix}  h_{\q\q} & 0\\ 0 & h_{\phi\phi}\end{pmatrix} = \begin{pmatrix}  \f{1}{1+ \f{\Theta^2}{\b^2} \sin^2 \q}& 0\\ 0 & \f{\sin^2 \q}{\left(1+ \f{\Theta^2}{\b^2} \sin^2 \q\right)^2}\end{pmatrix}.
\ee
The thermal effective action begins with the term
\be
-\log Z(\b, \Omega) =- c_0 \int d^2 x \sqrt{h} \f{1}{\b^2} = -\f{c_0}{\b^2} \int d^2 x \f{\sin \q}{\left(1+ \f{\Theta^2}{\b^2} \sin^2 \q\right)^{3/2}} = \f{-4\pi c_0}{\b^2\left(1-\Omega^2\right)}\,.
\ee
We can also work out the first correction from angular velocity. This is because the leading corrections to the thermal effective action are given by 
\be
-\log Z(\b, \Omega) = \int d^2 x \sqrt{h} \left(-c_0 \f{1}{\b^2} + c_1 R+ c_2 F^2+ \cdots\right),
\ee
where $F_{ij} = \p_i A_j - \p_j A_i$. We have
\be
F^2 = h^{ij} h^{kl} F_{ik} F_{jl} =h^{\q\q} h^{\phi\phi} ((\partial_\q A_\phi)^2 + (-\partial_\q A_\phi)^2)) \,.
\ee
Doing the integrals in the effective action gives the series up to this order as \cite{Benjamin:2023qsc}
\be
-\log Z(\b, \Omega) =\f{4\pi}{ (1-\Omega^2)}\left(-\f{c_0}{\b^2} + 2 c_1(1-\Omega^2) - \f{8c_2}{3} \Omega^2+\cdots\right).
\ee
%In this calculation the Ricci scalar for the metric $h_{ij}$ integrated against $\sqrt{h}$ gives $8\pi$.

\subsection*{$d=4$}
In this section we will consider spatial manifolds $S^3$ and $S^1 \times S^2$ with metrics
\be
ds^2 = d\t^2 + d\q^2 + \sin^2 \q \,d\phi_1^2 + \cos^2 \q\, d\phi_2^2\,,\qquad \q \in [0, \pi/2]
\ee
\vspace{-3mm}
\be
ds^2 = d\t^2 + d\q^2 + \sin^2 \q d\phi_1^2   + d\phi_2^2\,,\qquad \q \in [0, \pi]\,,
\ee
respectively. We can turn on angular velocities in the $\phi_1$ and $\phi_2$ directions. This means we have periodicity
\be
(\t, \phi_1, \phi_2) \sim (\t + \b, \phi_1 + \Theta_1, \phi_2 + \Theta_2)\,.
\ee
Since we need to put the metric in KK form \eqref{kkmetric} we again define
\be
\phi_i \longrightarrow \phi_1 + \f{\Theta_i}{\b} \t\,
\ee
and perform a Weyl transformation (ignoring the anomaly which is important in this 4d case; it will enter as an exponential correction to $Z$) to map our metrics to
\be
ds^2 = d\t^2 + \f{d\q^2 + \sin^2 \q \, d\phi_1^2 + \cos^2\q \, d\phi_2 ^2 + 2 \f{\Theta_1}{\b} \sin^2 \q \, d\t d\phi_1 + 2 \f{\Theta_2}{\b} \cos^2 \q \, d\t d\phi_2}{1+ \f{\Theta_1^2}{\b^2} \sin^2 \q + \f{\Theta_2^2}{\b^2}\cos^2 \q} 
\ee
and
\be
ds^2 = d\t^2 + \f{d\q^2  + \sin^2\q \, d\phi_1 ^2 + d\phi_2^2 + 2 \f{\Theta_1}{\b}  \sin^2 \q\,d\t d\phi_1 + 2 \f{\Theta_2}{\b}  \, d\t d\phi_2}{1+ \f{\Theta_1^2}{\b^2}\sin^2 \q + \f{\Theta_2^2}{\b^2}}\,.
\ee
In this case there is also an anomaly contribution since our spatial $S^1 \times S^2$ manifold does not give a 4d conformally flat manifold, but we focus here on just the contribution from the thermal effective action. Comparing to the KK metric and noticing that
\be
h_{ij} + A_i A_j = g_{ij}
\ee
leads to a spatial crossterm, we find  
\be
A_i = (A_\q, A_{\phi_1}, A_{\phi_2}) = \left(0, \f{\f{\Theta_1}{\b} \sin^2 \q}{1+ \f{\Theta_1^2}{\b^2}\sin^2 \q + \f{\Theta_2^2}{\b^2}\cos^2 \q}, \f{\f{\Theta_2}{\b} \cos^2 \q}{1+ \f{\Theta_1^2}{\b^2} \sin^2 \q+ \f{\Theta_2^2}{\b^2}\cos^2 \q}\right)
\ee
\be
\hspace{-3mm} \begin{pmatrix} h_{\q\q} & 0 & 0\\ 0 & h_{\phi_1 \phi_1} & h_{\phi_1 \phi_2} \\ 0 & h_{\phi_2 \phi_1} & h_{\phi_2\phi_2}
\end{pmatrix} = \begin{pmatrix} \f{1}{1+ \f{\Theta_1^2}{\b^2} \sin^2 \q+ \f{\Theta_2^2}{\b^2}\cos^2 \q} & 0 & 0\\ 0 & \f{\sin^2 \q \left(1+ \f{\Theta_2^2}{\b^2}\cos^2 \q\right)}{\left(1+ \f{\Theta_1^2}{\b^2} \sin^2 \q+ \f{\Theta_2^2}{\b^2}\cos^2 \q\right)^2} & -\f{\f{\Theta_1 \Theta_2}{\b^2} \sin^2 \q \cos^2 \q}{\left(1+ \f{\Theta_1^2}{\b^2} \sin^2 \q+ \f{\Theta_2^2}{\b^2}\cos^2 \q\right)^2} \\ 0 & -\f{ \f{\Theta_1 \Theta_2} {\b^2}\sin^2 \q \cos^2 \q}{\left(1+ \f{\Theta_1^2}{\b^2} \sin^2 \q+ \f{\Theta_2^2}{\b^2}\cos^2 \q\right)^2} &\f{\cos^2 \q \left(1+ \f{\Theta_1^2}{\b^2}\sin^2 \q\right) }{\left(1+ \f{\Theta_1^2}{\b^2} \sin^2 \q+ \f{\Theta_2^2}{\b^2}\cos^2 \q\right)^2}
\end{pmatrix}
\ee
for the $S^3$ case. The $S^1 \times S^2$ case has the same formulas as above except we set every factor of $\cos^2 \q$ to $1$. For the $S^3$ case we get
\be
\sqrt{h} = \f{\sin \q \cos \q}{\left(1 +\f{\Theta_1^2}{\b^2}\sin^2 \q+ \f{\Theta_2^2}{\b^2} \cos^2 \q\right)^2}
\ee
which is the same formula we got by boosting in \eqref{boost4d} so leads to the same universal correction. In the $S^1 \times S^2$ case we instead get
\be
\sqrt{h} = \f{\sin \q}{\left(1 +\f{\Theta_1^2}{\b^2}\sin^2 \q+ \f{\Theta_2^2}{\b^2} \right)^2}\,,
\ee
which integrates to (setting $\Theta_i^2/\b^2 = - \Omega_i^2$)
\be
-\log Z(\b, \Omega_1, \Omega_2) \approx -\f{c_0}{\b^3}\left(\frac{1}{\left(1-\Omega_2^2\right) \left(1-\Omega_1^2-\Omega_2^2\right)}+\frac{\tan ^{-1}\left(\frac{\Omega_1}{\sqrt{1-\Omega_1^2-\Omega_2^2}}\right)}{\Omega_1 \left(1-\Omega_1^2-\Omega_2^2\right)^{3/2}}\right)\,.
\ee
This gives the correct answer for $\Omega_1 = 0$, which should just be a boost in one direction for a toroidal black hole. %In the $S^3$ case we used $\int \sqrt{h} R = \f{4\pi^2 (3 - 2 \Omega_1^2 - \Omega_2^2)}{(1-\Omega_1^2)(1-\Omega_2^2)}$. 
\subsection*{General dimension}
As seen from the examples above, the EFT approach will always give the same integrals as the method using Lorentz invariance from Section \ref{spherelorentz}. The full effective action to order $\b^{3-d}$ for the thermal partition function on $S^1_\b \times S^{d-1}$ with angular velocities $\vec{\Omega}$ turned on is given by \cite{Benjamin:2023qsc}:
\be
\hspace{-16mm}-\log Z(\b, \vec{\Omega}) = \f{\text{Vol}(S^{d-1})}{\prod_{i}(1-\Omega_i^2)} \left(-\f{c_0}{\b^{d-1}} + \f{1}{\b^{d-3}}\left((d-2)(d-1)c_1 -\left(2c_1 + \f{8}{d}c_2\right)(d-2) \sum_{i}\Omega_i^2\right) +\cdots \right).
\ee

\small
\bibliographystyle{ourbst}
\bibliography{ModularSphere.bib}

\end{document}